\documentclass[aps,prl,twocolumn,superscriptaddress,showpacs,longbibliography,floatfix,nofootinbib]{revtex4}

\usepackage[utf8]{inputenc}
\usepackage{slashed}
\date{\today}

\usepackage{color}
\usepackage{graphicx}   
\usepackage{bm}
\usepackage{amsmath}
\usepackage{amsfonts}
\usepackage{hyperref}

\def\sumint{\,\hbox{$\sum$}\!\!\!\!\!\!\!\int}
\newcommand{\be}{\begin{equation}}
\newcommand{\ee}{\end{equation}}
\newcommand{\ba}{\begin{eqnarray}}
\newcommand{\ea}{\end{eqnarray}}

\begin{document}

\title{Shear Viscosity at Infinite Coupling: A Field Theory Calculation}

\author{Paul Romatschke}
\affiliation{Department of Physics, University of Colorado, Boulder, Colorado 80309, USA}
\affiliation{Center for Theory of Quantum Matter, University of Colorado, Boulder, Colorado 80309, USA}

\begin{abstract}
  I derive an exact integral expression for the ratio of  shear viscosity over entropy density $\frac{\eta}{s}$ for the massless (critical)  O(N) model at large N with quartic interactions. The calculation is set up and performed entirely from the field theory side using a non-perturbative resummation scheme that captures all contributions to leading order in large N. In 2+1d, $\frac{\eta}{s}$ is evaluated numerically at all values of the coupling. For infinite coupling, I find $\frac{\eta}{s}\simeq 0.42(1)\times N$. I show that this strong coupling result for the viscosity is universal for a large class of interacting bosonic O(N) models.
\end{abstract}

\maketitle

The shear viscosity is a transport coefficient that encodes how efficiently spatial anisotropies are transmuted into momentum anisotropies (velocity gradients). For relativistic systems, a key dimensionless ratio that quantifies this efficiency is found by the ratio of shear viscosity $\eta$ to entropy density $s$.
Experimental measurements of momentum anisotropies in heavy-ion collisions together with hydrodynamic modeling constrain the value of shear viscosity in QCD to $\frac{\eta}{s}\lesssim 0.2$ \cite{Romatschke:2007mq,Dusling:2007gi,Schenke:2010rr,Song:2013qma,Bernhard:2019bmu}.

This numerical value happens to be not too dissimilar from the result $\frac{\eta}{s}=\frac{1}{4\pi}\simeq 0.08$ found for the conjectured strong-coupling limit of another gauge theory, ${\cal N}=4$ Super-Yang-Mills (SYM) theory, in the large N limit \cite{Policastro:2001yc,Kats:2007mq,Buchel:2008sh}.
By contrast, in QCD where calculations are limited to weak coupling only, one typically encounters values $\frac{\eta}{s}\gg \frac{1}{4\pi}$ \cite{Arnold:2000dr,Arnold:2003zc,Ghiglieri:2018dib}.

At intermediate couplings, results for the shear viscosity exist for QED with many fermion flavors \cite{Moore:2001fga}
and for SU(3) gauge theory from lattice Monte-Carlo simulations \cite{Meyer:2007ic,Pasztor:2018yae}. At strong coupling,  results for transport coefficients have been limited to theories with known holographic duals, with the exception of so-called ``thermodynamic transport coefficients'' such as $\kappa$ \cite{Romatschke:2019gck}. In particular, there is no known example of a theory where $\frac{\eta}{s}$ has been determined for \textit{all coupling strengths}.

The present work is meant to fill this gap, and provide the complete coupling-dependence for the shear viscosity over entropy ratio by directly calculating the relevant low-frequency limit of the energy-momentum tensor correlator from the field theory.  The theory to be studied will be the O(N) vector model with quartic interactions in 2+1 dimensions in the large N limit, which exists for all values of the coupling. The choice of this theory is motivated by the fact that the entropy density $s$ is known for all couplings  \cite{Romatschke:2019ybu} and that an efficient resummation scheme that captures all relevant contributions to the shear viscosity at large N is known \cite{Romatschke:2019rjk,Romatschke:2019wxc}. It also helps that the most difficult part of the calculation, namely the evaluation of the shear viscosity coefficient for the O(N) vector model, has already been set up for the 3+1d theory using a variant of the 2PI formalism in Ref.~\cite{Aarts:2004sd}. Thus strictly speaking, the only new result in the present work will be the determination of $\frac{\eta}{s}$ at infinite coupling, which can be done in the 3d large N O(N) model (but does not make sense  because of the  Landau pole in 4d).

A major objection to calculating the shear viscosity in a theory with only two space dimensions is the presence of so-called long-time tails \cite{Forster:1976zz}, which normally lead to a divergent two-dimensional shear viscosity when naively taking the low frequency limit. However, for the O(N) model it so happens that $\frac{\eta}{s}\propto N$, so that in the large N limit long-time tails are suppressed by three powers of N, cf. the discussion in Ref.~\cite{Kovtun:2011np}. For this reason, the calculation of the shear viscosity as the zero-frequency limit of the relevant energy-momentum correlator reported in this work is well defined.

\textbf{Preliminaries}

In the interest of brevity, I will not review the set up of finite temperature correlators in quantum field theory (the interested reader is referred to the Supplemental Material for this topic). Denoting Minkowski momenta as ${\cal K}=(\omega,{\bf k})$ and space-time coordinates as ${\cal X}=(t,{\bf x})$, I will use the relation of retarded real-time Greens functions $G_R$ and the spectral density $\rho$ given by
\be
\label{spectralGR}
G_R({\cal K})=-\int \frac{d\mu}{\pi}  \frac{ \rho(\mu,{\bf k})}{\mu-\omega-i 0^+}\,.
\ee
This relation can be used to derive the analytic continuation of the corresponding Euclidean correlator to real time (cf. Refs.~\cite{Kovtun:2012rj,Romatschke:2017ejr})
\be
\label{anaki}
G_R(\omega)=-G_E(\omega_E\rightarrow i \omega-0^+)\,.
\ee

In order to connect the retarded correlator with transport coefficients, one employs the low-momentum expansion of $G_R$ provided by fluid dynamics. Fluid dynamics is the effective field theory of conserved quantities for small moment. For a theory that conserves energy and momentum, good EFT variables are the energy density and fluid 4-velocity $\epsilon,u^\mu$, fulfilling $u^\mu u_\mu=-1$ (see e.g. Ref.~\cite{Kovtun:2012rj,Florkowski:2017olj,Romatschke:2017ejr} for reviews of modern fluid dynamics). Using fluid dynamics, it is straightforward to derive the relation $G_R(\omega,{\bf k}=0)=P-i\omega \eta+{\cal O}(\omega^2)$ for the $T^{xy}$ correlator in $d\geq 3$ spacetime dimensions, from which the so-called Kubo relation follows:
\be
\label{Kubo}
\lim_{\omega\rightarrow 0}\frac{\partial}{\partial {\omega}}\rho^{xy,xy}(\omega,{\bf k}=0)=\eta\,.
\ee
Including thermal fluctuations in the fluid dynamic calculations leads to a long-time tail contribution for $d=3$ of the form $G_R(\omega,{\bf k}=0)\propto \frac{i \omega T^2}{\eta/s}\ln \omega$ (see Supplemental Material). This term in general invalidates the Kubo formula (\ref{Kubo}), except in the large N limit where it is suppressed by $\frac{\eta}{s}\propto N$, thereby allowing the use of (\ref{Kubo}) to extract the shear-viscosity for $d=3$ to ${\cal O}(N^0)$.

\textbf{The O(N) model}

The field theory I consider in this work is that of a massless (``critical'') N-component scalar field $\phi_a$, $a=1,2,\ldots, N$ with Euclidean action
\be
\label{action}
S=\int d^{d}X \left[\frac{1}{2}\partial_\mu \phi_a \partial_\mu \phi_a+\frac{\lambda}{N}\left(\phi_a\phi_a\right)^2\right]\,,
\ee
where at finite temperature $T$ the $X^0=\tau$ direction is compactified on a circle with radius $\beta$. The partition function for this theory $Z=\int {\cal D}\phi e^{-S}$ may be rewritten by inserting $1=\int {\cal D}\sigma \delta(\sigma-\phi_a\phi_a)=\int {\cal D}\sigma {\cal D}\zeta e^{i \int \zeta (\sigma-\phi_a\phi_a)}$. Integrating out the $\sigma$ field gives
\be
\label{pathintegral}
Z=\int {\cal D}\phi {\cal D}\zeta e^{-\int_X\left[\frac{1}{2}\partial_\mu \phi_a \partial_\mu \phi_a+i \zeta \phi_a\phi_a +\frac{N}{4\lambda}\zeta^2\right]}\,.
\ee
Splitting the field $\zeta=\frac{1}{2}\zeta_0+\zeta^\prime$ into a zero mode and fluctuations, the action in (\ref{pathintegral}) becomes
\ba
\label{r0res}
S&=&S_{R0,0}+S_{R0,I}+\frac{N V \beta}{16\lambda}\zeta_0^2\,,\\
S_{R0,0}&=&\frac{1}{2}\int_X\left[\partial_\mu \phi_a \partial_\mu \phi_a+i \zeta_0 \phi_a\phi_a+\frac{N}{2\lambda}\zeta^{\prime 2}\right]\,,\nonumber
\ea
where $S_{R0,I}=i \int_X \zeta^\prime \phi_a\phi_a$.

At leading order in large N, only the zero-mode from the $\zeta$-field contributes to the partition function, so $S_{R0,I}$ may be ignored. This corresponds to a particular resummation of finite-temperature Feynman diagrams (certain tadpoles), dubbed the R0-level resummation in Ref.~\cite{Romatschke:2019rjk}. One finds for R0
\be
\label{zr0}
Z_{R0}=\sqrt{\frac{N V \beta}{16\pi\lambda}}\int d\zeta_0 e^{-\frac{N V \beta}{16\lambda} \zeta_0^2-N V \beta J(\sqrt{i \zeta_0})}\,,
\ee
where $V$ is the $d-1$ dimensional volume of space and $e^{-N V \beta J(\sqrt{i \zeta_0})}\equiv\int {\cal D}\phi e^{-\frac{1}{2}\int_X\left[\partial_\mu \phi_a \partial_\mu \phi_a+i \zeta_0\phi_a\phi_a\right]}$. In the large N limit, the remaining ordinary integral in (\ref{zr0}) can be evaluated exactly from the saddle point located at $i \zeta_0=m^2$, so that $Z_{R0}=e^{-N V \beta\left(m^4-J(m)\right)}$, where
\be
J(m)=\frac{1}{2}\int \frac{d^{d-1}{\bf k}}{(2\pi)^d}\left[E_k+2 T \ln \left(1-e^{-E_k \beta}\right)\right]\,,
\ee
$E_k=\sqrt{{\bf k}^2+m^2}$ \cite[Eq.~(2.44)]{Laine:2016hma}. Using dimensional regularization for $d=3-0^+$, the saddle point condition becomes
\be
\label{saddle}
\frac{\pi}{\hat{\lambda}} \hat{m}^2+\hat{m}+2 \ln\left(1-e^{-\hat{m}}\right)=0\,,
\ee
where $\hat{m}\equiv \beta m$, $\hat{\lambda}\equiv \beta \lambda$. Note that in the strong coupling limit $\hat{\lambda} \rightarrow \infty$ (\ref{saddle}) has a universal solution $\hat{m}=2\ln \frac{1+\sqrt{5}}{2}$ \cite{Romatschke:2019ybu,Sachdev:1993pr}. Basic thermodynamic relations give the exact large N entropy density as $s=\frac{\partial}{\partial T}\frac{\ln Z_{R0}}{\beta V}$. Using (\ref{saddle}), integration by parts, and a variable change to $E=E_k$, for d=3 this relation leads to
\ba
\label{sr0}
s&=&-\frac{N\beta}{4\pi} \int_{m}^\infty dE E (E^2-m^2) n^\prime(E)\,,\\
&=&\frac{N T^2}{2\pi}\left[2 \hat{m}^2\ln n(m)+6 \hat{m} {\rm Li}_2\left(e^{-\hat{m}}\right)+6 {\rm Li}_3\left(e^{-\hat{m}}\right)\right]\,,\nonumber
\ea
where $n(x)=\left(e^{\beta x}-1\right)^{-1}$, $n^\prime(x)=\partial_x n(x)$.

For some correlation functions, additional interactions not included in the R0 resummation may contribute at leading order in large N, making it necessary to go beyond the R0-level. To this end, slightly changing notation from Ref.~\cite{Romatschke:2019rjk} rewrite the action (\ref{r0res}) as $S_{R0,0}+S_{R0,I}=S_{R2,0}+S_{R2,I}$ with
\ba
\label{r2res}
S_{R2,0}&=&\frac{1}{2}\int_{X,Y}\left[\phi_a \Delta^{-1}\phi_a+\zeta^\prime D^{-1}\zeta^\prime\right]\,,\\
S_{R2,I}&=&i \int_X \zeta^\prime \phi_a\phi_a-\frac{1}{2}\int_{X,Y}\left[\phi_a \Sigma\phi_a+\zeta^\prime N \Pi \zeta^\prime \right]\,,\nonumber
\ea
where
\be
\label{propagators}
\Delta(K)=\frac{1}{K^2+m^2+\Sigma(K)}\,,\quad
D(K)=\frac{1}{N}\frac{1}{\frac{1}{2\lambda}+\Pi(K)}\,.
\ee
The R2-level resummation then consists of calculating $\Sigma,\Pi$ self-consistently up to (including) one-loop level using $S_{R2,I}$, finding
\be
\label{r2self}
\Sigma_{R2}(X)=2 D(X) \Delta(X)\,,\quad \Pi_{R2}(X)=4 \Delta^2(X)\,.
\ee
At large $N$, $D\propto \frac{1}{N}$, so $\Sigma$ is generically subleading, while $\Pi$ is not. 

At finite temperature, more subtle ways generating contributions at leading order in large N arise, making it necessary to go beyond the R2 resummation. To this end, rewrite the action (\ref{r0res}) again using $S_{R2,0},S_{R2,I}$ but express
\be
\label{r3vertex}
i \int_X \zeta^\prime \phi_a\phi_a=
+i\int_{X,Y,Z}\zeta^\prime \phi_a\phi_a\left[\Gamma_3-\delta \Gamma_3\right]\,,
\ee
where $\Gamma_3(P,K)=1+\delta \Gamma_3(P,K)$ is the resummed three-vertex function and $P,K$ are the incoming momenta for the $\Delta$-propagators. The R3 level resummation then consists of calculating $\Gamma_3$ self-consistently to one-loop level, and $\Sigma,\Pi$ self-consistently up to (including) two-loop level. Diagrammatically, one finds \cite{Romatschke:2019rjk}
\be
\label{selfr3}
\delta \Gamma_3=-4\includegraphics[trim=-5 25 0 0, width=0.12\linewidth]{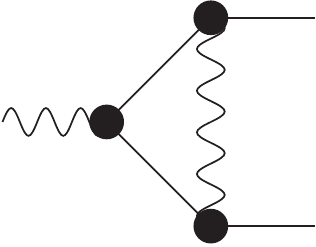}\,,\quad
\Pi=2\includegraphics[trim=-5 20 0 0 ,width=0.14\linewidth]{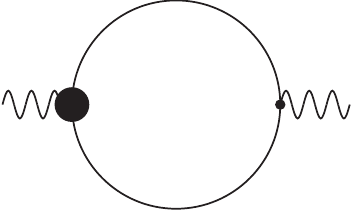}\,,\quad
\Sigma=4\includegraphics[trim= -5 0 0 0, width=0.14\linewidth]{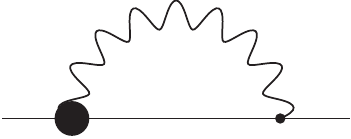}
\ee
where the full vertex is denoted by $$\Gamma_3=\includegraphics[trim=0 25 0 0, width=0.12\linewidth]{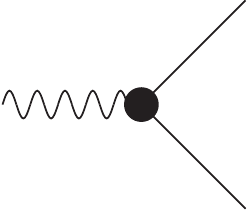}\,,$$ wiggly lines denote $D(K)$ and straight lines denote $\Delta(K)$.

While the R0 resummation contains all leading order large N contributions for the zero-point function, R2 and R3 contain all large N contributions for the two- and three-point function, respectively. The energy-momentum tensor is a four-point function, so including all large N contributions requires going to R4. The R4 resummation is found by adding and subtracting a term $\int_{X,Y,U,Z} \phi_a \phi_a \phi_b \phi_b \Gamma_4$. Calculating the one-loop 4-point vertex self-consistently, one finds diagrammatically
\be
\label{vertex4}
\Gamma_4=-2\includegraphics[trim=0 40 0 0,width=0.15\linewidth]{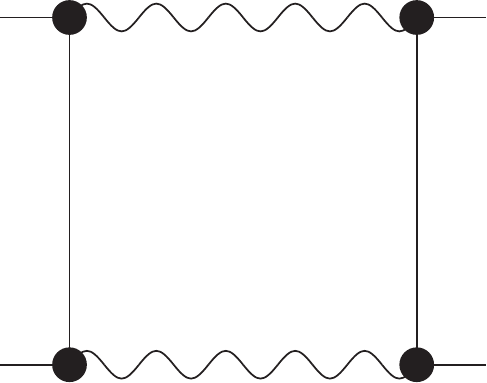}\,,
\ee
whereas the large N three-point vertex in R4 becomes
\be
\label{r4vertex}
\delta \Gamma_3=-4\includegraphics[trim=-5 25 0 0, width=0.12\linewidth]{vertex2}-8 N \includegraphics[trim=-5 15 0 0, width=0.3\linewidth]{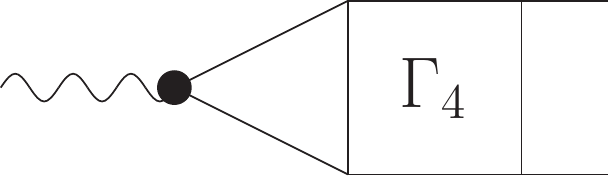}\,,
\ee
and the self-energies $\Pi,\Sigma$ are still diagrammatically given by (\ref{selfr3}). Note that since $D\propto N^{-1}$, the 4-point vertex and the triangle diagram contribute at the same order at large N.

\textbf{Energy-Momentum Tensor Correlators}

For the action (\ref{action}), the Euclidean operator for the energy-momentum tensor is given by $T^{xy}(X)=\partial_x \phi_a(X) \partial_y \phi_a(X)$ so that the Euclidean energy-momentum tensor correlator is defined by
\be
G_E^{xy,xy}(X)=\frac{\int {\cal D}\phi e^{-S}T^{xy}(X)T^{xy}(0)}{Z}\,.
\ee
In the R0-level resummation, following (\ref{r0res}) and neglecting $S_{R0,I}$, the R0-action is quadratic in $\phi$ and hence $G_E^{xy,xy}$ can be calculated using Wick's theorem. At finite temperature, $\phi_a(X)=T\sum_{k_E}\int_{{\bf k}}e^{i K\cdot X}\phi_a(K)$ where $k_E=2 \pi n T$ are the bosonic Matsubara frequencies with $n\in \mathbb{Z}$. In Fourier space with ${\bf p}=p \hat{x}$, one finds \cite{Romatschke:2019ybu,Romatschke:2019gck}
\be
\label{ger0}
G_{E,R0}^{xy,xy}(P)=2 N \sumint_K \frac{(k_x-p)^2k_y^2}{(K^2+m^2)((K-P)^2+m^2)}\,,
\ee
where the mass $m$ is determined by (\ref{saddle}). The result may be analytically continued to real frequencies $\omega$ using Eq. (\ref{anaki}). The R0-level result $G_{R,R0}^{xy,xy}(\omega,{\bf p}=0)$ constitutes the correct large N result for the retarded correlator \textit{except for the region} $\beta \omega \ll \frac{1}{N}$.

To appreciate this statement, let us reconsider $G_E^{xy,xy}$ in the R2-level resummation. Using (\ref{r2res}) and neglecting $S_{R2,I}$, the R2-action is once again quadratic in the fields, and hence $G_E^{xy,xy}$ can be calculated using Wick's theorem. Setting again ${\bf p}\rightarrow 0$ gives
\be
\label{ger2}
G_{E,R2}^{xy,xy}(p_E)=2 N \sumint_K k_x^2k_y^2 \Delta(K)\Delta(P-K)\,,
\ee
with $\Sigma=\Sigma_{R2}$ in (\ref{propagators}) and $\Sigma_{R2}$ specified by (\ref{r2self}). Using the spectral representation of the propagator
\be
\label{rhose}
\rho({\cal K})=-\frac{{\rm Im}\,\Sigma({\cal K})}{\left({\cal K}^2+m^2+{\rm Re}\,\Sigma({\cal K})\right)^2+\left({\rm Im}\,\Sigma({\cal K})\right)^2}\,,
\ee
the thermal sum and angular integral is straightforward, giving for $d=3$
\ba
\label{ger2v2}
G_{E,R2}^{xy,xy}(p_E)&=&\frac{N}{4} \int_0^\infty\frac{dk k^5}{2\pi} 
\int \frac{d \mu d\mu^\prime}{2\pi^2} \frac{\rho(\mu)\rho(\mu^\prime)}{\mu+\mu^\prime+i p_E}\nonumber\\
&&\times\left(\coth\frac{\mu}{2 T}+\coth\frac{\mu^\prime}{2T}\right)\,,
\ea
%
Note that since $\Sigma\propto \frac{1}{N}$, in the naive large N limit (\ref{rhose}) becomes the spectral function for a free massive particle, $\rho(\omega)=\pi {\rm sign}(\omega)\delta\left({\cal K}^2+m^2\right)$.
Continuing $G_E^{xy,xy}$ analytically to real frequencies $\omega_E\rightarrow i \omega-0^+$, the imaginary part becomes
\ba
\label{s2123}
\rho^{xy,xy}_{R2}(\omega)&=&\frac{N}{2} \int_{\cal K} {\bf k}^4
\rho(k^0)\rho(\omega-k^0)
\left(n(k^0)-n(k^0-\omega)\right)\,.\qquad
\ea
One finds that the product  $\rho(k^0)\rho(k^0-\omega)$ has
contributions where poles ``pinch'' the real $k^0$ axis from above and below in the large N limit. Focusing on the limit $\beta \omega\ll \frac{1}{N}$, this
contribution becomes
\be
\label{rep}
\rho^2(k^0)\rightarrow \frac{1}{2}\Delta_R(k^0)\Delta_A(k^0)=-\frac{\rho(k^0)}{2{\rm Im}\,\Sigma(k^0,{\bf k})}\,,
\ee
which is proportional to $N$ in the large N limit. Note that the other contributions (as well as the whole product $\rho(k^0)\rho(k^0-\omega)$ for $\beta \omega \gg \frac{1}{N}$) give contributions of order ${\cal O}(N^0)$, 
%
%
bringing us back to the R0 result (\ref{ger0}).
The non-commutative nature of the limits $\beta \omega\rightarrow 0$, $\frac{1}{N}\rightarrow 0$ imply that for the calculation of transport coefficients, contributions that are naively subleading at large N become important. This enhancement process, known as ``pinching poles'', has a long history in nuclear physics literature, cf. Refs.~\cite{Jeon:1994if,Aarts:2004sd}.

In the small frequency limit, the R2-level expression for the shear viscosity from Eq.~(\ref{Kubo}) is 
\ba
\label{myeta}
\eta_{R2}&=&\frac{N}{16\pi}\int_m^\infty dE \left.\frac{k^4 n^\prime(E)}{{\rm Im}\, \Sigma(E)}\right|_{k=\sqrt{E^2-m^2}}\,.
\ea

The R2-level expression for the shear viscosity is well-defined for all values of the coupling, but it does not contain all contributions to leading order in large N. To this end, let us reconsider the correlator $G_E^{xy,xy}$ in R4-level resummation, specified by (\ref{r2res}) with (\ref{r4vertex}). Since the R4-level action contains a non-trivial vertex, Wick's theorem can no longer be used to evaluate the correlator; instead, interactions must be included (See the Supplemental Material for an example of how a generic 4-point function is evaluated beyond R2.) For the energy-momentum correlator $G_{E,R4}^{xy,xy}$, the situation is slightly simpler than for a generic 4-point function, because the spatial derivatives $\partial_x \phi, \partial_y \phi$ in the definition imply that some contributions do not contribute  after angular integration
(see Supplemental Material). However, instead of the regular 3-vertex $\Gamma_3$, the corresponding contribution to $G_{E,R4}^{xy,xy}$ contains momenta $k_x,k_y$ inside the vertex. One thus finds
\be
G_{E,R4}^{xy,xy}(p_E)=2 N \sumint_K k_x k_y \Gamma_{3,xy}(K,P-K)\Delta(K)\Delta(P-K)\,,
\ee
where $\Gamma_{3,xy}=k_x k_y+\delta \Gamma_{3,xy}$ denotes the resummed 3-vertex, and the propagator $\Delta(K)$ contains $\Sigma$ in the R4-level resummation, cf. Eq.~(\ref{selfr3}).  The resummed vertex is the only modification with respect to the R2 result (\ref{ger2}), so one needs to check if $\delta \Gamma_{3,xy}$, which naively is order $\frac{1}{N}$ gets enhanced in the limit $P\rightarrow 0$. One finds
\ba
\label{vertex1}
\delta \Gamma_{3,xy}(K,P-K)&=&-4 \sumint_Q \Delta(Q)\Delta(P-Q)W(P,Q,K)\nonumber\\
&&\hspace*{-3cm}\times \Gamma_3(-Q,P-K)\Gamma_{3}(K,Q-P)\Gamma_{3,xy}(Q,P-Q)\,,
\ea
where the integral kernel is $W(P,Q,K)=D(P-K-Q)+2 N \Gamma_4(Q,P-Q,K,P-K)$. 
The structure of (\ref{vertex1}) indeed suggests a ``pinching pole'' similar to (\ref{ger2}) in the limit $P\rightarrow 0$, whereas for other kinematic regions $\delta \Gamma_{3,xy}\propto \frac{1}{N}$. Two of the internal vertices in (\ref{vertex1}) therefore do not receive corrections in the limit $P\rightarrow 0$. One may verify that in this limit, $\delta \Gamma_{3,xy}(K,-K)=\frac{k_x k_y}{k^2} \bar \Gamma_3(K)$, such that
after doing the angular average one obtains
\be
\label{ger4}
G_{E,R4}^{xy,xy}(p_E)=\frac{N}{4} \sumint_K k^2 \bar\Gamma_{3}(K,P-K)\Delta(K)\Delta(P-K)\,,
\ee
where to leading order in large N
\ba
\label{vertex1v2}
\bar\Gamma_{3}(K,P-K)&=&k^2-4 \sumint_Q \Delta(Q)\Delta(P-Q)W(P,Q,K)\nonumber\\
&&\times  \bar \Gamma_{3}(Q,P-Q)\times \left(2 ({\bf \hat{q}}\cdot{\bf \hat{k}})^2-1\right)\,.
\ea

The analytic structure of the vertex $\bar \Gamma_3(K,P-K)$ is as follows: expressing the propagators in terms of their spectral functions $\rho(\mu),\rho_D(\mu)$, one can perform the analytic continuation to real frequencies. In a first step, setting $\bar \Gamma_3(Q,P-Q)=q^2$, the thermal sum over $q_E$ in (\ref{vertex1}) can be done explicitly. Since the integrations over the arguments of the spectral functions range over the whole real axis, one finds that $\bar\Gamma_3(i k^0,ip^0-ik^0)$ has branch cuts along the whole real line for ${\rm Im}\,p^0=0$, ${\rm Im}\,k^0=0$, ${\rm Im}\,(p^0-k^0)=0$. This structure is unchanged when iterating the vertex.

With the analytic structure of the vertex known, one proceeds to evaluate (\ref{ger4}). First, write the thermal sum as
\be
\label{ts}
T\sum_{k_E} f(k_E)=\oint_{\cal C} \frac{dk^0}{4\pi i} \coth\frac{\beta k^0}{2} f(i k^0)\,,
\ee
with ${\cal C}$ encircling the poles at the imaginary Matsubara frequencies $k^0=i k_E$ in a counter-clockwise fashion. Next, the propagators $\Delta(K)\Delta(P-K)$ have branch cuts at ${\rm Im}\,k^0=0, {\rm Im}(k^0)=-p_E$. The additional branch cut from the vertex is independent from $k^0$, and hence not relevant for the thermal sum (\ref{ts}). Deforming the contour ${\cal C}$ such that it runs along both sides of the two branch cuts, one encounters terms such as
\be
\bar\Gamma_3\left(i k^0\pm 0^+,p_E-i k^0\right) \Delta\left(i k^0\pm 0^+\right) \Delta\left(p_E- i k^0 \right)\,.
\ee
Upon analytic continuation $p_E\rightarrow i p^0-0^+$, one recognizes the retarded and advanced correlators $\Delta_{R,A}$ using (\ref{spectralGR}), where $\Delta_R(-k^0)=\Delta_A(k^0)$. The leading large N contributions are then given by combinations such as $\Delta_R(k^0)\Delta_A(k^0-\omega)$ which have singularities on either side of the real $k^0$ axis (``pinching poles''), whereas the others can be neglected. Letting $k^0\rightarrow k^0-p^0$ in the second branch cut contribution, only a particular analytic continuation of the vertex function contributes, which for $p^0\rightarrow 0$ and $k^0=\pm E_k$ becomes
\be
\label{Gareg}
\lim_{p^0\rightarrow 0}\left.\bar\Gamma_3(ik^0-0^+,i(p^0-k^0)- 0^+,{\bf k})\right|_{k^0=\pm E_k}\equiv F(E_k)\,.
\ee
Clearly, this corresponds to use standard analytic continuation of $\bar\Gamma_3(K,P-K)$ \textit{first} for $k_E$ and \textit{then} for $p_E$.
%
%
%
%
%
Assuming that $F(E_k)$ is real (which will be shown below),  one can then follow the same procedure that led to (\ref{myeta}), with the only modification arising from the resummed vertex function.
%
Using (\ref{sr0}), one finds
\be
\label{fulleta}
\frac{\eta}{s}=\frac{1}{4}\frac{\int_m^\infty dE \frac{F(E)}{\beta {\rm Im}\, \Sigma(E)}k^2 n^\prime(E)}{-\int_m^\infty dE E k^2 n^\prime(E)}\,,
\ee
where again $k=\sqrt{E^2-m^2}$.
This relation is exact in the large N limit, as it contains all leading order large N contributions to the shear viscosity and entropy density, cf. Ref~\cite{Aarts:2004sd}. Note that because $\Sigma\propto \frac{1}{N}$, one finds that $\frac{\eta}{s}\propto N$ in the large N limit. This N-scaling is generic for vector or fermionic theories \cite{Aarts:2004sd,Moore:2001fga}, but is qualitatively different from large N gauge theories where $\eta/s\propto {\cal O}\left(N^0\right)$ \cite{Arnold:2000dr,Policastro:2001yc}

\begin{figure*}[t]
  \includegraphics{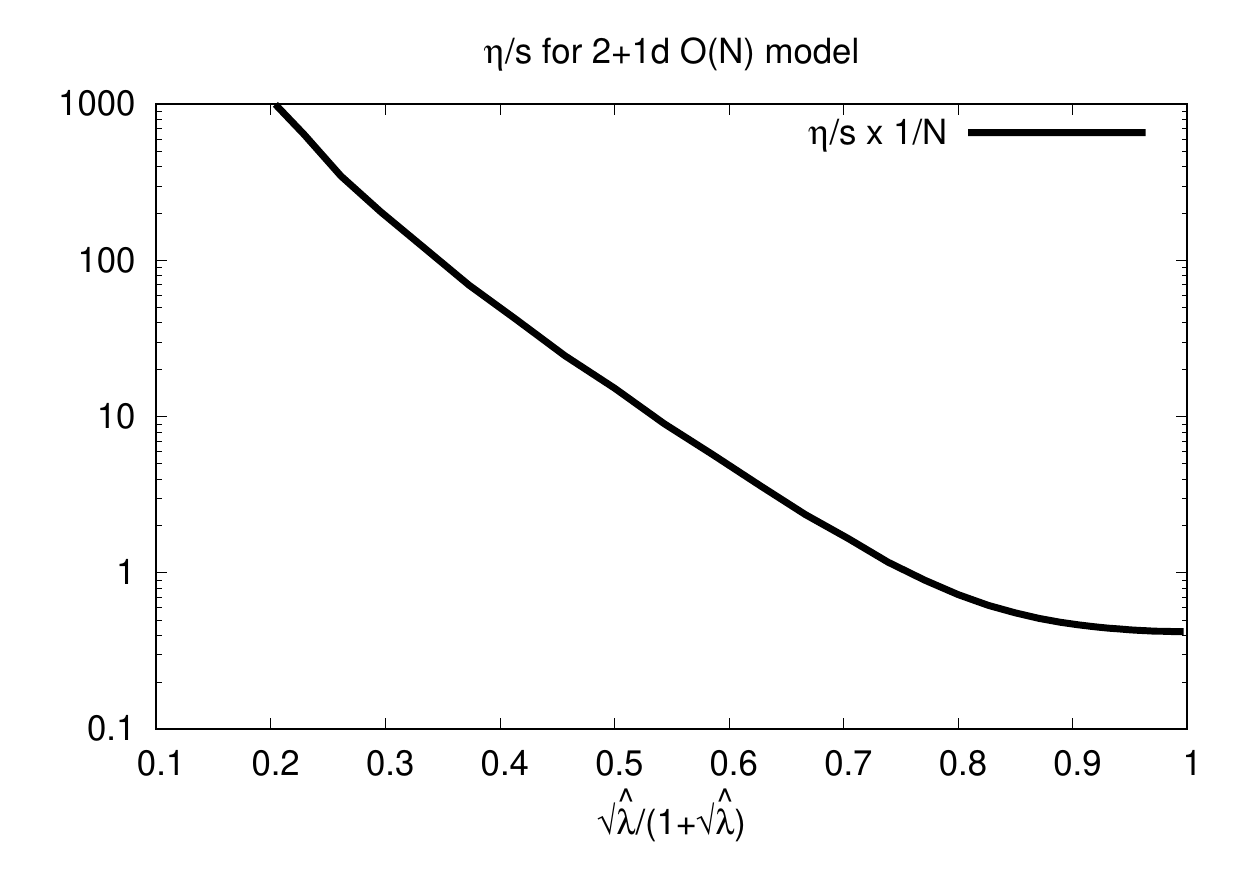}
  \caption{\label{fig1} Shear viscosity over entropy density as a function of the dimensionless coupling $\hat{\lambda}\equiv \beta \lambda$ in the 2+1d O(N) model with quartic self-interaction. Horizontal axis is compactified in order to fit values $\hat{\lambda} \in [0,\infty)$. The numerical code to calculate this result is publicly available from Ref.~\cite{codedown}.}
    \end{figure*}

In order to get a result for $\frac{\eta}{s}$, one needs to know the functions ${\rm Im}\,\Sigma(E)$ and $F(E)$. These follow from finite-temperature field theory calculations, see Supplemental Material and Refs.~\cite{ValleBasagoiti:2002ir,Aarts:2004sd}. For the evaluation of $\frac{\eta}{s}$, it is convenient to use quadrature to recast the integrals in terms of sums (see Supplemental Material). To this end, construct orthogonal polynomials $P_n(x)$ of degree $n$
\be
\label{ortho}
-\int_m^\infty dE\, n^\prime(E) P_n(E) P_m(E)=\delta_{n,m}\,,
\ee
for $n=0,1,\ldots, K$. Expanding 
$\frac{F(E_k)}{{\rm Im}\, \Sigma(E_k)}=\sum_{n=0}^\infty b_n P_n(E_k)$
only the coefficients $b_0,b_1,b_2$ contribute to $\frac{\eta}{s}$ because $k^2$ in the numerator of (\ref{fulleta}) only involves polynomials $P_n(E)$ up to degree two.

\textbf{Results and Universality at Infinite Coupling}

Numerical evaluation of (\ref{fulleta}) for all values of the dimensionless coupling $\hat{\lambda}=\beta \lambda$ is shown in Fig.~\ref{fig1}. For weak couplings $\hat{\lambda}$, $\frac{\eta}{s N}$ is large because the thermal width is small. As the coupling is increased, I find that the ratio of shear viscosity over entropy density is dropping monotonically, but is finite in the limit of $\hat{\lambda}\rightarrow \infty$. In this strong coupling limit, the numerically calculated result becomes
\be
\label{etares}
\lim_{\hat{\lambda}\rightarrow \infty} \frac{\eta}{s}=0.42(1) \times N\,,
\ee

One may ask about the universality of the strong coupling result. To this end, consider a modification of the  action (\ref{action}) to 
\be
\label{action2}
S=\int d^{d}X \left[\frac{1}{2}\partial_\mu \phi_a \partial_\mu \phi_a+\frac{\lambda}{N^2}\left(\phi_a\phi_a\right)^3\right]\,,
\ee
where now $\lambda$ is dimensionless. This action has the property that it is a CFT for all values of $\lambda$ at large N \cite{Romatschke:2019ybu}. Using the same replacement $\sigma=\phi_a\phi_a$ as before, and integrating out $\sigma$ one finds
\be
\label{phi6}
Z=\int {\cal D}\phi {\cal D}\zeta e^{-\int_X\left[\frac{1}{2}\partial_\mu \phi_a \partial_\mu \phi_a+i \zeta \phi_a\phi_a-\ln\left({\rm Ai}\left(i \zeta \left(\frac{N^2}{3 \lambda}\right)^{\frac{1}{3}}\right)\right)\right]}\,.
\ee
At large N, the asymptotic properties of the Airy function then give the form of the partition function as
\be
\label{phi62}
Z=\int {\cal D}\phi {\cal D}\zeta e^{-\int_X\left[\frac{1}{2}\partial_\mu \phi_a \partial_\mu \phi_a+i \zeta \phi_a\phi_a+\frac{2}{3 \sqrt{3\lambda}} N (i \zeta)^{\frac{3}{2}}\right]}\,.
\ee
While it may be challenging to construct R4 for this Lagrangian for general values of $\lambda$, the strong coupling limit $\lambda\rightarrow \infty$ of this theory is exactly equal to the strong coupling limit $\hat{\lambda}\rightarrow \infty$ of Eq.~(\ref{pathintegral}). For this reason, one can explicitly write down R4 for this theory in the strong coupling limit where the only difference is the form of the propagator
\be
D(K)=\frac{1}{N \Pi(K)}\,,
\ee
cf. Eq.~\ref{r2res}. Hence \textit{in the infinite coupling limit} the result for the shear viscosity for (\ref{action2}) is identical to that of (\ref{action}). It is not hard to generalize this proof to theories with other interactions $\phi_a\phi_a\rightarrow U(\phi_a\phi_a)$ for a large class of potentials $U(x)$, which demonstrates that(\ref{etares}) is the universal strong coupling shear viscosity over entropy ratio for a large class of bosonic QFTs.
The same universal behavior was found \cite{Romatschke:2019ybu} for the weak-strong ratio $\frac{s_{\lambda=\infty}}{s_{\lambda=0}}=\frac{4}{5}$ and for the boson in-medium mass $\lim_{\lambda\rightarrow \infty} \hat{m}=2 \ln \frac{1+\sqrt{5}}{2}$.

\textbf{Summary and Discussion}

In this work, I derived an exact large N expression for the ratio of shear viscosity over entropy density for the interacting O(N) model, using well-established field theory techniques. Evaluating the expression numerically in the case of 2+1 dimensions, I found the strong coupling result (\ref{etares}).

Regardless of the numerical value, the present work demonstrates that it is possible to calculate transport properties at infinite coupling directly from quantum field theory, without invoking dualities or conjectures of any kind. While the field theory studied here may not be of interest to most readers, it can nevertheless serve as a test bed for strong coupling transport which would otherwise be inaccessible or very hard by any other means. For instance, having access to exact full energy-momentum tensor correlation functions  for all values of the coupling allows to study the onset/breakdown of hydrodynamics from first principles, cf. Refs.~\cite{Romatschke:2015gic,Grozdanov:2016vgg,Heller:2020jif}; exact real-time correlators also can be used to test analytic continuation techniques employed in lattice Monte-Carlo studies \cite{Meyer:2007ic,Pasztor:2018yae};  exact results for transport coefficients can be used as a rigorous test case for approximation schemes that are used for e.g. QCD \cite{Arnold:2000dr,Arnold:2003zc,Ghiglieri:2018dib}.

In addition to serving as a well-defined test bed for general-purpose tools, the present calculation may be generalized in several ways. By, for instance, calculating other transport coefficients such as the bulk viscosity $\zeta$ as well as relaxation time $\tau_\pi$ for the O(N) model; calculating transport coefficients for other large N theories \cite{DeWolfe:2019etx,Pinto:2020nip,Romatschke:2019qbx}; calculating exact far-from equilibrium real-time dynamics at large N for a quantum field theory.

For these reasons, I am optimistic that the present result can become useful in the future. 

\textbf{Acknowledgments}

 I am indebted to Gert Aarts for clarifying some questions I had concerning Ref.~\cite{Aarts:2004sd}, as well as providing numerical data for $\frac{\eta}{s}$ in the 4d O(N) model from this reference. Also, I thank Scott Lawrence and Max Weiner for fruitful discussions, and Marcus Pinto for pointing out a typo in (\ref{phi62}).  This work was supported by the Department of Energy, DOE award No DE-SC0017905.


\onecolumngrid

\newpage

\section{Supplemental Material}

This supplemental material contains details about the calculation as well as examples not presented in the main text.

\subsection{Finite Temperature Correlators}

Given the operator $\hat{\phi}({\cal X})$, one can define the Wightman functions
$G_{>}({\cal X})=\langle \hat\phi({\cal X})\hat\phi(0)\rangle$ and $G_{<}({\cal X})=\langle \hat\phi(0)\hat\phi({\cal X})\rangle$, which in turn
are related to the retarded real-time Greens function \cite[Eq.~(2.9)]{Kovtun:2012rj}
\be
\label{gr}
G_{R}(t,{\bf x})=-i\theta(t)\langle \left[\hat\phi({\cal X}),\hat\phi(0)\right]\rangle
\ee
via $G_{R}(t,{\bf x})=-i \theta(t)\left[G_>({\cal X})-G_<({\cal X})\right]$.
As shown in \cite[Eq.~(8.10)]{Laine:2016hma}, for thermal equilibrium, the Wightman functions fulfill the KMS relation $G_>(t-i \beta)=G_<(t)$, where $\beta=\frac{1}{T}$ is the inverse temperature. 
Using mostly plus metric convention, the Fourier transform $G({\cal K})=\int_{\cal X} e^{-i {\cal X}\cdot {\cal K}} G({\cal X})$ of
the KMS relation becomes $G_>({\cal K}) e^{-\beta \omega}=G_<({\cal K})$. The spectral function is defined as
\be
\label{spectral}
\rho({\cal K})=\frac{G_>({\cal K})-G_<({\cal K})}{2}=\frac{G_>({\cal K})\left(1-e^{-\beta \omega}\right)}{2}\,.
\ee
In thermal equilibrium, all correlators can be related to each other, and can be expressed through the spectral function. In particular,
using  the representation of the step function $\theta(t)=i \int \frac{d \nu}{2\pi} \frac{e^{-i \nu t}}{\nu+i 0^+}$ one proves the relation
\be
\label{spectralGR2}
G_R({\cal K})=-\int \frac{d\mu}{2\pi}  \frac{2\rho(\mu,{\bf k})}{\mu-\omega-i 0^+}\,,
\ee
which is called the spectral representation of the retarded correlator. Using the definition of the advanced correlator $G_A({\cal K})=G_R^*({\cal K})$, this relation can be used to show
\be
\label{spfo}
\rho({\cal K})=-{\rm Im}\, G_R({\cal K})=\frac{i}{2}\left(G_R({\cal K})-G_A({\cal K})\right)\,.
\ee

One can relate these real-time Greens function to the imaginary-time Greens function obtained in the Euclidean formulation. Euclidean momenta and coordinates will be denoted
by $K=\left(k_E,{\bf k},k_4\right)$ and $X=\left(\tau,{\bf x}\right)$, respectively. 
Defining the Euclidean correlator for imaginary time $\tau$ as
\be
G_E(\tau,{\bf x})=\langle \phi(X)\phi(0)\rangle_E\,,
\ee
this matches $G_>(t,{\bf x})$ if  formally identifying $\tau\rightarrow i t$. Therefore,
\be
\label{ana1}
G_>(t,{\bf x})\equiv G_E(\tau=i t,{\bf x})\,,\quad G_E(\tau,{\bf x})=G_>(t=-i\tau,{\bf x})\,.
\ee
Using these relations, and using (\ref{spectral}), one may then prove 
\be
\label{Espec}
G_E(K)=\int \frac{d\mu}{2\pi}  \frac{2 \rho(\mu,{\bf k})}{\mu+ik_E}\,,
\ee
which is called the spectral representation of the Euclidean correlator. Comparing (\ref{Espec}), (\ref{spectralGR2}), the retarded correlator for real frequencies $\omega$ is obtained by analytic continuation of the Euclidean frequencies (\ref{anaki}),
such that $\rho({\cal P})={\rm Im}\, G_E(\omega_E\rightarrow i \omega-0^+,{\bf p})$.

\subsection{Fluid Dynamics in two space  dimensions}

In fluid dynamics, using the EFT variables $\epsilon, u^\mu$ as building blocks, one constructs the energy-momentum tensor in a gradient expansion as
\be
\label{gradexp}
T^{\mu\nu}=\epsilon u^\mu u^\nu+p g^{\mu\nu}-\eta \sigma^{\mu\nu}-\zeta \Delta^{\mu\nu}+\ldots\,,
\ee
where $g_{\mu\nu}$ is the metric tensor, $p=p(\epsilon)$ is the pressure related to the energy density via the equilibrium equation of state, $\eta,\zeta$ are the shear and bulk viscosity coefficients, respectively, $\Delta^{\mu\nu}=g^{\mu\nu}+u^\mu u^\nu$ and $\sigma^{\mu\nu}=\nabla_\perp^{\mu} u^{\nu}+\nabla_\perp^{\nu} u^{\mu}-\frac{2}{d-1}\Delta^{\mu\nu} \nabla^\perp_\lambda u^\lambda$, $\nabla_\perp^\mu=\Delta^{\mu\nu}\partial_\mu$ and $d$ denotes the number of space-time dimensions, cf. Ref.~\cite[Eq.~(2.30)]{Romatschke:2017ejr}. Higher-order versions of fluid dynamics exist, but will not be considered here. In order to calculate the real time correlator of the energy-momentum tensor, one considers the linear response of $T^{\mu\nu}$ with respect to metric fluctuations $\delta g_{\mu\nu}$ around Minkowski space,
\be
G_R^{\alpha\beta,\gamma \delta}({\cal X})=-2\left.\frac{\delta T^{\alpha\beta}}{\delta g_{\gamma\delta}}\right|_{g={\rm Minkowski}} \,.
\ee
Fourier transforming, one finds \cite[Eq.~(2.101)]{Romatschke:2017ejr}
\ba
\label{gr0i0j}
G_R^{0i,0j}({\cal K})=(\epsilon+P)\left[\frac{k^i k^j}{{\bf k}^2} \frac{\omega^2}{\omega^2-c_s^2k^2+i \omega k^2 \gamma_s}
  +\left(\delta^{ij}-\frac{k^i k^j}{{\bf k}^2}\right)\frac{k^2 \gamma_\eta}{i\omega-\gamma_\eta k^2}\right]\,,
\ea
where $\gamma_\eta=\frac{\eta}{\epsilon+P}$, $\gamma_s=\frac{\frac{2 (d-2)}{(d-1)}\eta+\zeta}{\epsilon+P}$ and $c_s^2=\frac{dP}{d\epsilon}$ is the speed of sound squared. Using $\partial_\mu T^{\mu\nu}=0$, neglecting contact terms and letting ${\bf k}=k \hat{x}$, this leads to
\be
\label{grclassical}
G_R^{xx,xx}=\frac{(\epsilon+P)\omega^2(c_s^2-i \omega \gamma_s)}{\omega^2-c_s^2 k^2+i \omega k^2 \gamma_s}\,,\quad
G_R^{xy,xy}=\frac{(\epsilon+P) \omega^2 \gamma_\eta}{i\omega-\gamma_\eta k^2}\,.
\ee
In particular, find for the spectral functions for $d=3$ and $k=0$ one finds
\be
\label{Kubo-sup}
\rho^{xx,xx}(\omega)=\left(\eta+\zeta\right)\omega\,,\quad
  \rho^{xy,xy}(\omega)=\eta \omega\,.
\ee

The poles of the retarded correlator (\ref{gr0i0j}) show that the fluid dynamics EFT contains collective modes (sound modes, shear modes). These modes will interact, thereby modifying the form of (\ref{gradexp}). It is possible to estimate the effect of these interactions for low momenta by calculating the contribution to $G_R$ arising from hydrodynamic self-interactions. Since this is an effective theory calculation, the resulting momentum integrals have to be cut-off at a scale $p_{\rm max}\propto \gamma_\eta^{-1}$ below which the EFT can be trusted. Specifically, focusing on the shear mode contribution, one finds for $d=3$  \cite{Kovtun:2012rj}
\be
\label{grshearshear}
G_{\rm R,1-loop,shear-shear}^{xy,xy}(\omega,{\bf k}=0)=\frac{1}{8\pi}\int_0^{p_{\rm max}} dp \frac{-p^3}{p^2-\frac{i \omega}{2 \gamma_\eta}}={\rm const.}-\frac{i \omega}{32 \pi \gamma_\eta}\ln \frac{i p_{\rm max}^2}{\frac{ \omega}{2\gamma_\eta}}\,,
\ee
which needs to be added to $G_R^{xy,xy}$ in Eq.~(\ref{grclassical}). Additional contributions to $G_R^{xy,xy}$ result from the sound-sound mode contribution (same order as (\ref{grshearshear})), as well as higher-order loops (suppressed by powers of the temperature).

\subsection{Example: 4-point correlator in R2, R3}

As an example for the use of the R-level resummations, consider the zero-temperature 4-point function
\be
\label{ce}
C(X)=\langle \phi_i^2(X)\phi_j^2(0)\rangle\,.
\ee
In R2, recognizing $\phi_i^2(X)=\sigma(X)$, integrating out the $\sigma$ field as with the partition function (\ref{pathintegral}) leads to $
C(X)=\frac{N}{2\lambda}\delta^d(X)-\frac{N^2}{4\lambda^2}D(X)$ or in Fourier space
\be
\label{cp2}
C(P)=\frac{N \Pi(P)}{1+2 \lambda \Pi(P)}\,.
\ee
At weak coupling $\hat{\lambda}\ll 1$, one may expand $D(X)$ in a power series, finding $C(X)= 2 N \Delta^2(X)$. For $d=3$, the zero-temperature propagator is proportional to $\Delta(X)\propto \frac{1}{X}$, so that $C(X)\propto \frac{N}{X^2}$ for $\lambda\ll 1$. Similarly, $\Pi_{R2}(X)\propto \frac{1}{X^2}$ so that for strong coupling $\lim_{\hat{\lambda}\rightarrow \infty} D(X)\propto \frac{1}{N X^4}$. Neglecting the contact term, this leads to 
$
C(X)\propto \frac{N}{\lambda^2 X^4}$ for $\hat{\lambda} \gg 1$. 
which explicitly proves the statement that the dimension of the operator changes from the free theory limit (``UV fix point'') to the strong coupling limit (``IR fix point''), cf. \cite{Klebanov:2002ja}.

It is instructive to reconsider this correlator at higher R-level resummation. At R3, one has
\ba
\label{cex}
C&=&2 N\includegraphics[trim=0 25 0 0, width=0.06\linewidth]{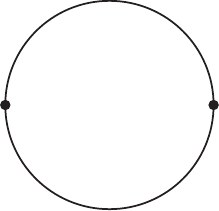}-8 N
\includegraphics[trim=0 25 0 0,width=0.06\linewidth]{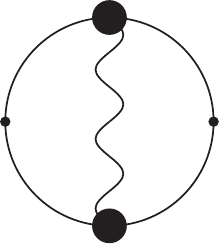}
+32 N \includegraphics[trim=0 25 0 0,width=0.06\linewidth]{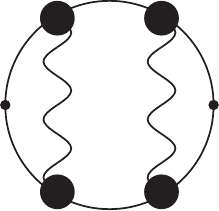}+\ldots
-4 N^2 \includegraphics[trim=0 25 0 0 ,width=0.18\linewidth]{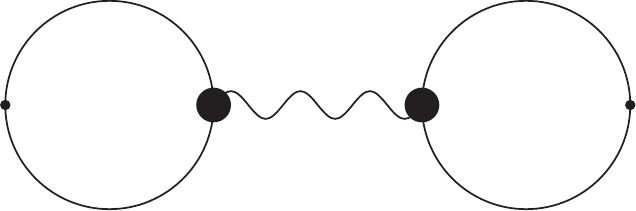}\,,
\ea
as well as many other contributions that mutually cancel in R3, such as
\be
4 N^2 \includegraphics[trim=0 25 0 0,width=0.18\linewidth]{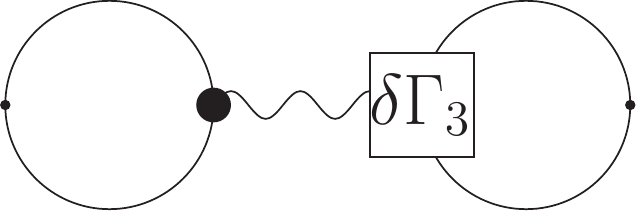}
+16 N^2\includegraphics[trim=0 25 0 0,width=0.18\linewidth]{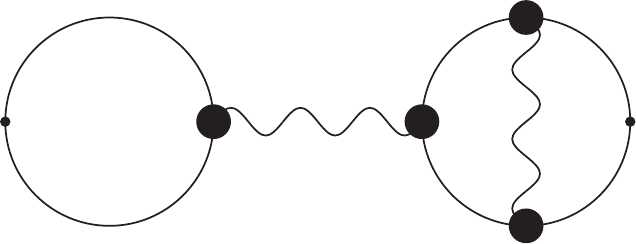}=0\,,\nonumber
\ee
or
\ba
2 N \includegraphics[trim=0 25 0 0,width=0.06\linewidth]{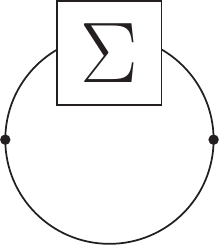}
-8 N \includegraphics[trim=0 25 0 0,width=0.06\linewidth]{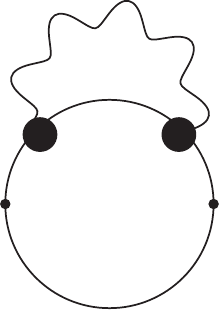}
+16 N\includegraphics[trim=0 25 0 0, width=0.06\linewidth]{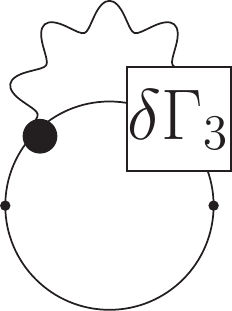}
+32 N \includegraphics[trim=0 25 0 0, width=0.06\linewidth]{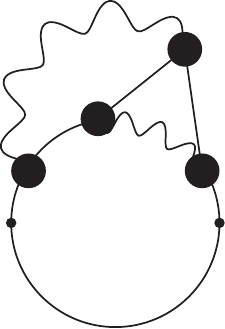}=0\,,\nonumber
\ea
where boxes indicate insertions of self-energies or vertex functions in the R3-level resummation.

In addition, one finds that the first line in (\ref{cex}) can be summed-up in closed form using the resummed vertex $\Gamma_3$, finding
\be
\label{cr3}
C=2 N \includegraphics[trim= 0 25 0 0,width=0.06\linewidth]{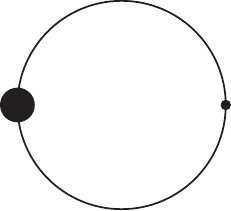}-4 N^2 \includegraphics[trim= 0 25 0 0,width=0.18\linewidth]{barbell}=\frac{N \Pi(P)}{1+2 \lambda \Pi(P)}\,,
\ee
with the only difference to the R2 result (\ref{cp2}) being that the $\Pi$ is to be evaluated in the R3-resummation, cf. (\ref{selfr3}).

\subsection{Thermal width}

From (\ref{selfr3}), the self-energy $\Sigma$ is given by
\be
\Sigma(P)=4\sumint_K \Delta(K) D(P-K) \Gamma_3(P,-K)\,.
\ee
The vertex function is ${\cal O}\left(\frac{1}{N}\right)$ for generic momenta, with a pinching-pole enhancement only when $K=P$. Since the corresponding integration region has measure $\frac{1}{N}$, the resummed vertex does not contribute to $\Sigma(P)$ to leading order in large N.

Using the spectral representation of the propagators, performing the thermal sum is straightforward. Upon analytic continuation $p_E\rightarrow i p^0-i 0^+$, one has
\ba
   {\rm Im}\Sigma({\cal P})&=&4 \int_{\cal K} \rho_D({\cal P}-{\cal K}) \rho({\cal K})
   \left(\coth\frac{\beta k^0}{2}+\coth\frac{\beta (p^0-k^0)}{2}\right)
\,.\nonumber
\ea
This expression can be further simplified by using the large N free particle result for $\rho({\cal K})$ and $\coth\frac{\beta x}{2}=1+2n(x)$ to find
\ba
\label{imsigma}
   {\rm Im}\Sigma({\cal P})&=& \int_{\bf k}\frac{2}{E_k} \left[\rho_D(p^0-E_k,{\bf p}-{\bf k})
     \left(n(E_k)-n(E_k-p^0)\right)
     +\rho_D(p^0+E_k,{\bf p}-{\bf k})
     \left(n(E_k)-n(E_k+p^0)\right)\right]
\,.\quad
\ea
Since only the on-shell limit will be needed, I denote ${\rm Im}\, \Sigma(E_p)\equiv {\rm Im}\, \Sigma(p^0=E_p,{\bf p})$, and  define the angular averages
\be
\label{avi}
\bar{\rho}_D(E_p\pm E_k)=\int\frac{d\phi}{2\pi} \rho_D(E_p\pm E_k,{\bf p}-{\bf k})\,,
\ee
so that
\be
\label{imsigma2}
      {\rm Im}\Sigma(E_p)=\frac{1}{\pi}\int_m^\infty dE_k \left[\bar{\rho}_D(E_p-E_k)
     \left(n(E_k)-n(E_k-E_p)\right)+\bar{\rho}_D(E_p+E_k)
     \left(n(E_k)-n(E_k+E_p)\right)\right]\,.
%
%
\ee
Note that while $n(E_k-E_p)$ has a pole at $E_k=E_p$, $\bar{\rho}_D(E_p-E_k)$ has a zero there, so that the pole is integrable.

\subsection{Polarization Tensor}

To evaluate (\ref{imsigma2}), the spectral function for the auxiliary propagator $\rho_D({\cal K})$ is needed. This in turn depends on the polarization tensor (\ref{selfr3})
\be
\label{pi}
\Pi(P)=2 \sumint_K \Delta(P-K)\Delta(K)\Gamma_3(K,P-K)\,.
\ee
This expression has pinching poles and an order ${\cal O}(N^0)$ vertex correction for $P\rightarrow 0$. However, the corresponding contribution to (\ref{imsigma}) has vanishing integral weight in the large N limit, so to leading order in N one may neglect both the resummed vertex as well as the pinching poles in (\ref{pi}). As a consequence, the required contribution for the polarization tensor is given by the simple sum-integral
\be
\Pi(P)=2 \sumint_K \frac{1}{K^2+m^2}\frac{1}{(P-K)^2+m^2}\,,
\ee
with $m$ given by the solution from (\ref{saddle}). I write the integral as two parts $\Pi(P)=\Pi_{V}(P)+\Pi_T(P)$ where $\Pi_T(P)$ contains the Bose-Einstein distribution factor after evaluating the thermal sum, while $\Pi_V$ does not. For $\Pi_V=2 \int_K \Delta(P-K)\Delta(K)$ one finds
\ba
\Pi_V(P)
&=&\frac{i}{4\pi \sqrt{P^2}} \ln \frac{1-\frac{i \sqrt{P^2}}{2m}}{1+\frac{i \sqrt{P^2}}{2m}}\,. 
\ea
This may be analytically continued to give
\ba
\Pi_V({\cal P})&=&\theta\left({\cal P}^2\right) \frac{\arctan\frac{\sqrt{{\cal P}^2}}{2m}}{2\pi \sqrt{{\cal P}^2}}
+\theta\left(-{\cal P}^2\right)\frac{1}{4 \pi \sqrt{-{\cal P}^2}}\ln\left|\frac{2m+\sqrt{-{\cal P}^2}}{2m-\sqrt{-{\cal P}^2}}\right|
+\theta\left(-{\cal P}^2-4m^2\right) {\rm sign}(p^0)\frac{i}{4 \sqrt{-{\cal P}^2}}\,,
\nonumber
\ea
where ${\cal P}^2=-(p^0)^2+{\bf p}^2$. For $\Pi_T(P)$ one finds
\ba
\Pi_T(P)&=&4 \int_{\bf k}
  \frac{n(E_k)}{E_k} \frac{P^2-2{\bf p}\cdot {\bf k}}{E_1^4-2 E_1^2(E_2^2-p_E^2)+(E_2^2+p_E^2)^2}\nonumber\,,
\ea
where $E_1=E_k$, $E_2=E_{\bf p-k}$ and I have used the symmetry ${\bf k}\rightarrow {\bf p-k}$ to simplify the integrand. Further simplification yields
\be
  \Pi_T(P)=4 \int_{\bf k}
  \frac{n(E_k)}{E_k} \frac{P^2-2{\bf p}\cdot {\bf k}}{(P^2-2 {\bf p}\cdot {\bf k})^2+4 p_E^2 E_k^2}\nonumber\,.
  \ee
 Taking apart the integrand, one can use \cite[Eq.~(B6)]{Romatschke:2003ms} to integrate over angles and 
 after analytically continuing to real frequencies  I find 
\ba
   {\rm Re}\,\Pi_{T}({\cal P})&=&-\frac{1}{\pi}\int_m^\infty dE\, n(E)
   \left[\frac{{\rm sign}\left(-{\cal P}^2+2 p^0 E\right)\theta\left(-{\cal P}^2\left(p^0+2 E\right)^2+p^2 (4 m^2+{\cal P}^2)\right)}{\sqrt{-{\cal P}^2\left(p^0+2 E\right)^2+p^2 (4 m^2+{\cal P}^2)
     }}+
     \right.\nonumber\\
     &&\left.+\frac{{\rm sign}\left(-{\cal P}^2-2 p^0 E\right)\theta\left(
-{\cal P}^2\left(p^0-2 E\right)^2+p^2 (4 m^2+{\cal P}^2)
\right)}{\sqrt{
-{\cal P}^2\left(p^0-2 E\right)^2+p^2 (4 m^2+{\cal P}^2)
       }}\right]\,,\\
        {\rm Im}\,\Pi_{T}({\cal P})&=&-\frac{1}{\pi}\int_m^\infty dE\, n(E)\left[{\rm sign}(-p^0+E)\frac{\theta\left(-p^2({\cal P}^2+4m^2)+{\cal P}^2\left(p^0-2 E\right)^2\right)}{\sqrt{-p^2({\cal P}^2+4m^2)+{\cal P}^2\left(p^0-2 E\right)^2}}+\right.\nonumber\\
          &&\left.+
          {\rm sign}(-p^0-E)\frac{\theta\left(-p^2({\cal P}^2+4m^2)+{\cal P}^2\left(p^0+2 E\right)^2\right)}{\sqrt{-
p^2({\cal P}^2+4m^2)+{\cal P}^2\left(p^0+2 E\right)^2
          }}
          \right]\nonumber\,.
        \ea
        The remaining integrals can be done numerically, where it is useful to employ integration by parts. Using $\Pi({\cal P})=\Pi_V({\cal P})+\Pi_T({\cal P})$ the spectral function for the auxiliary correlator is
        \be
        \rho_D({\cal P})=\frac{1}{N}{\rm Im}\,\frac{1}{\frac{1}{2\lambda}+\Pi({\cal P})}=-\frac{1}{N} \frac{{\rm Im}\,\Pi({\cal P})}{\left(\frac{1}{2\lambda}+{\rm Re}\,\Pi({\cal P})\right)^2+\left({\rm Im}\,\Pi({\cal P})\right)^2}\,.
        \ee
Since ${\rm Im}({\cal P})=0$ for ${\cal P}^2\in[-4m^2,0]$, the spectral function is also vanishing in this region.

        \subsection{Vertex Resummation -- R3}

The relevant part of Eq.~(\ref{vertex1}) is the pinching-pole contribution to the three-vertex. For pedagogical reasons, let me first evaluate this contribution in the R3-level resummation, where the integral kernel is $W(P,Q,K)=D(P-K-Q)$. One finds
        \be
        \label{vertex2}
        \bar\Gamma_{3}(K,P-K)=k^2-4 \sumint_Q \Delta(Q)\Delta(P-Q)D(P-K-Q)\bar\Gamma_3(Q,P-Q) f\left({\bf \hat{q}}\cdot{\bf \hat{k}}\right)\,,
        \ee
        with the angular weight
        \be
        \label{2dweight}
        f(x)=2 x^2-1
        \ee
        For (\ref{ger4}), only the case ${\bf p}=0$ is needed.  
        Using the spectral representation for $D(P-K-Q)=\int \frac{d\mu}{\pi}
        \frac{\rho_D(\mu)}{\mu+i(p_E-k_E-q_E)}$, the relevant thermal sum can be written as (\ref{ts})
        \be
        \label{asum}
        T\sum_{q_E} \frac{\Delta(Q)\Delta(P-Q)\Gamma_3(Q,P-Q)}{\mu+i(p_E-k_E-q_E)}
        =\oint \frac{dq^0}{4\pi i} \coth\frac{\beta q^0}{2}
        \frac{\Delta(iq^0)\Delta(p_E-i q^0) \Gamma_3(i q^0,p_E-i q^0)}{\mu+i(p_E-k_E)+q^0}\,,
        \ee
        where all the spatial arguments of the propagators and vertex are ${\bf q}$. 
        The contour is encircling the poles of $\coth\frac{\beta q^0}{2}$ on the imaginary axis. Apart from these poles, the integrand possesses a pole at $q^0=-\mu-i(p_E-k_E)$, a branch cut at ${\rm Im}\,q^0=0$ from $\Delta(iq^0)$ and $\Gamma_3(i q^0,p_E+i q^0)$, and a branch cut at ${\rm Im}\,q^0=-p_E$ from $\Delta(p_E-iq^0)$ and $\Gamma_3(i q^0,p_E-i q^0)$. The pole contribution to (\ref{asum}) is
        \be
        {\rm pole}=\frac{1}{2}\coth\frac{\beta \mu}{2} \Delta(-i\mu+p_E-k_E)\Delta(i\mu+k_E)\Gamma_3(-i\mu+p_E-k_E,i\mu+k_E)\,.
        \ee
        According to (\ref{Gareg}), first the analytic continuation for $k_E$ and then for $p_E$ must be done, leading to
        \be
           \lim_{p^0\rightarrow 0}({\rm pole})=\frac{1}{2}\coth\frac{\beta (q^0-k^0)}{2} \left.\Delta_R(q^0)\Delta_A(q^0)F(E_q)\right|_{q^0=\mu+k^0}\,.
        \ee
        The branch cut contributions can be calculated in complete analogy to the steps leading up to (\ref{fulleta}), finding \cite{ValleBasagoiti:2002ir}
        \be
        \lim_{p^0\rightarrow 0}({\rm branch})=-\frac{1}{2}\coth\frac{\beta q^0}{2} \left.\Delta_R(q^0)\Delta_A(q^0)F(E_q)\right|_{q^0=\mu+k^0}\,.
        \ee
        Both contributions are manifestly real, proving the above statement that $F(E_q)$ is real. Using (\ref{rep}), (\ref{avi}),  after integrating this gives
\be
        \label{vertexui}
        F(E_k)=k^2+\frac{1}{\pi}\int_m^\infty dE
        \frac{F(E_q)}{ {\rm Im}\, \Sigma(E_q)}
        \left[\bar{\bar{\rho}}_D(E_k-E_q)
     \left(n(E_q)-n(E_q-E_k)\right)-\bar{\bar{\rho}}_D(E_q+E_k)
     \left(n(E_q)-n(E_q+E_k)\right)\right]\,.
        \ee
        Note the typo in \cite[Eq.~(103)]{Aarts:2004sd}. 
        Here $\bar{\bar{\rho}}_D$ denotes the angular average with weight (\ref{2dweight}),
        \be
        \bar{\bar{\rho}}_D(E_p\pm E_k)=\int\frac{d\phi}{2\pi} \rho_D(E_p\pm E_k,{\bf p}-{\bf k}) f\left({\bf \hat{q}}\cdot {\bf \hat{k}}\right)
        \ee
        cf. Eqns.~(\ref{avi}), (\ref{imsigma2}). Using the orthogonal polynomials defined for the quadrature (\ref{ortho}), it is convenient to parametrize
        \be
        \label{expansion}
        \frac{k^2}{{\rm Im}\, \Sigma(k)}=\sum_{i=0}^K a_n P_n(E_k)\,,\quad
        \frac{F(E_q)}{{\rm Im}\, \Sigma(q)}=\sum_{i=0}^K b_n P_n(E_q)\,,
        \ee
        so that (\ref{vertexui}) has the solution $b_n=A_{nm}^{-1} a_m$ where
        \be
        A_{nm}=\delta_{nm}-\frac{1}{\pi}\int_m^\infty dx \int_m^\infty dy \frac{P_n(x) P_m(y)}{{\rm Im}\, \Sigma(x)} n^\prime(x)\left[\left(n(y)-n(y-x)\right)\bar{\bar{\rho}}_D(x-y)-\left(n(y)-n(y+x)\right)\bar{\bar{\rho}}_D(x+y)\right]\,.
        \ee
The values of $A_{nm}$ are calculated numerically.

\subsection{Vertex Resummation -- R4}

In the R4 resummation, one needs the 4-vertex (\ref{vertex4}) in the resummation of the three vertex (\ref{r4vertex}). Writing the three vertex (\ref{r4vertex}) using a similar momentum convention as in (\ref{vertex2}), one has 
\be
\label{vresr4}
\bar\Gamma_3(K,P-K)=k^2-4\sumint_Q \Delta(Q)\Delta(P-Q)\bar\Gamma_3(Q,P-Q) W(P,Q,K)f\left({\bf \hat{q}}\cdot {\bf \hat{k}}\right)\,,
\ee
where $W(P,Q,K)=D(P-K-Q)+2 N \Gamma_4(Q,P-Q,K,P-K)$.
In the large N limit, the four-vertex is given from (\ref{vertex4}) as
\be
\Gamma_4(Q,P-Q,K,P-K)=-2 \sumint_R \Delta(R-Q) D(R-P) \Delta(R-K) D(R)\,.
\ee
There are no pinching poles in this expression, so the thermal width in the propagators $\Delta$ can be neglected to leading order in large N.
Using (\ref{ts}) to calculate thermal sum, one gets contributions from each of the propagators in the loop. These can be calculated by using (\ref{Espec}) sequentially for each propagator, finding

\ba
\label{g4-2}
\Gamma_4&=&-2\int_{\cal R} \coth\frac{\beta r^0}{2} \rho_D({\cal R})\left[
  D(R-P)\Delta(R-K)\Delta(R-Q)+ D(R+P)\Delta(R+P-K)\Delta(R+P-Q)\right]
\\
&&-2\int_{\cal R} \coth\frac{\beta r^0}{2} \rho({\cal R}) \left[D(K+R)D(K+R-P) \Delta(R+K-Q)
  +D(Q+R)D(Q+R-P) \Delta(R+Q-K)\right]\,,
\nonumber
\ea
where $r_E=i r^0$. This expression is to be evaluated inside the resummation for the three vertex (\ref{vresr4}). The propagators $\Delta(Q)\Delta(P-Q)$ and the vertex $\Gamma_3(Q,P-Q)$ have branch cuts for ${\rm Im}q^0=0$, ${\rm Im}q^0=-p_E$. In addition to these, the thermal sum over $q_E$ also picks up the singularities of $\Gamma_4$, which from (\ref{g4-2}). These are singularities on the $q^0$ real axis (for $\Delta(R-Q)$, $D(Q+R)$), on the line ${\rm Im}q^0=-p_E$ (for $\Delta(R+P-Q)$, $D(Q+R-P)$), as well as on the line ${\rm Im}q^0=-k_E$ (for $\Delta(R+K-Q)$, $\Delta(R+K-Q)$). For the branch cut at ${\rm Im}q^0=0$, there is a pinching pole contribution to $\Gamma_3(K,P-K)$ which arises when analytically continuing $q_E\rightarrow i q^0-0^+$. After performing the $q^0$ integration, the analytic continuation for $k_E,p_E$ are fixed by (\ref{Gareg}). So one finds that for the branch cut at ${\rm Im}q^0=0$, $\Gamma_4$ must be analytically continued by taking
\be
\label{ac1}
q_E\rightarrow i q^0-0^+\,,\quad
k_E\rightarrow i k^0-0^+\,,\quad
p_E\rightarrow i p^0-0^+\,,
\ee
in this sequence. For the branch cut at ${\rm Im}q^0=-p_E$, the pinching pole contribution to $\Gamma_3(K,P-K)$ arises when analytically continuing $q_E\rightarrow p_E+i q^0+0^+$, and contributes with a minus sign. Finally, the contribution for ${\rm Im}q^0=-k_E$ are simple poles which also give rise to a pinching pole contribution to $\Gamma_3(K,P-K)$. Adding the two analytic continuations of $\Gamma_4$ for the branch cuts, one finds that the terms proportional to $\rho_D$ in (\ref{g4-2}) cancel (matching the observation in Ref.~\cite{Aarts:2004sd}). Needing only the $p^0\rightarrow 0$ limit for the viscosity, one obtains
\be
\sum_{\rm branch-contributions} 2N \Gamma_4=-16 N i\int_{\cal R} \left(n(r^0-k^0)-n(r^0-q^0)\right) \rho({\cal R-K}) \rho({\cal R-Q}) |D_R({\cal R})|^2\,,
\ee
where $D_R({\cal R})$ is the retarded $D$ correlator. This is to be contrasted with the branch-cut contributions for $D(P-K-Q)$ from (\ref{vresr4}), which is $-2 i \rho_D({\cal Q-K})$.  In addition to the branch-cut contribution, there is the pole contribution arising from $\Delta(R+K-Q), \Delta(R+K-Q)$.
After shifting $q^0\rightarrow q^0-k^0$, the pole contribution gives
\be
\sum_{\rm pole-contribution} 2N \Gamma_4=16 N i\int_{\cal R} \left(n(r^0-k^0)-n(r^0-q^0)\right) \rho({\cal R-K}) \rho({\cal R-Q}) |D_R({\cal R})|^2\,,
\ee
whereas the corresponding pole contribution from $D(P-K-Q)$ is $2i\rho_D({\cal Q-K})$.

All told, the three vertex contribution (\ref{vresr4}) thus becomes
\ba
F(k^0)&=&k^2+8\int_{\cal Q} \frac{F(E_q)}{{\rm Im}\Sigma({\cal Q})}\left(n(q^0-k^0)-n(q^0)\right)\rho({\cal Q})\Lambda({\cal Q,K})f\left({\bf \hat{q}}\cdot{\bf \hat{k}}\right)\,,\nonumber\\
\Lambda({\cal Q,K})&=&\rho_D({\cal Q-K})+8 N \int_{\cal R} \left(n(r^0-k^0)-n(r^0-q^0)\right) \rho({\cal R-K}) \rho({\cal R-Q}) |D_R({\cal R})|^2\,.
\ea
One can now do the angular integrations that are part of $\int_{\cal Q},\int_{\cal R}$, respectively. Writing
\be
f\left({\bf \hat{q}}\cdot{\bf \hat{k}}\right)=f\left({\bf \hat{q}}\cdot{\bf \hat{r}}\right)f\left({\bf \hat{k}}\cdot{\bf \hat{r}}\right)+\sin(2 {\bf \hat{q}}\cdot{\bf \hat{r}})\sin(2 {\bf \hat{k}}\cdot{\bf \hat{r}})\,,
\ee
the only non-vanishing angular integrals are of the form
\be
\bar{\bar{\rho}}({\cal R-K})\equiv\int\frac{d\phi_r}{2\pi} \rho({\cal R-K})f\left({\bf \hat{k}}\cdot{\bf \hat{r}}\right) ={\rm sign}(r^0-k^0)f(\xi)\frac{\theta\left(1-\xi^2\right)}{2 r k \sqrt{1-\xi^2}} \,,\quad \xi=\frac{m^2+r^2+k^2-(r^0-k^0)^2}{2 r k}\,.
\ee
Therefore, the angular average of $\Lambda({\cal Q,K})$ becomes
\be
\bar{\Lambda}({\cal Q,K})=\bar{\bar{\rho}}_D({\cal Q-K})+8 N \int_{-\infty}^\infty \frac{dr^0}{2\pi}\int_0^\infty\frac{d r r}{2\pi}\left(n(r^0-k^0)-n(r^0-q^0)\right) \bar{\bar{\rho}}({\cal R-K}) \bar{\bar{\rho}}({\cal R-Q}) |D_R({\cal R})|^2\,.
\ee
The corresponding expression may be evaluated numerically by performing the same expansion as in (\ref{expansion}).

\subsection{Quadrature}

This section provides some detail on how to construct the orthogonal polynomials $P_n(x)$ needed for Eq.~(\ref{ortho}). Finding the roots $x_n$ from $P_{K+1}(x_n)=0$ allows to recast any integral 
\be
\label{quadrature}
\int_m^\infty dx\, n^\prime(x) f(x)=\sum_{i=0}^{K} w_i f(x_i)\,,
\ee
with $f(x)$ a polynomial of degree $2K+1$ or lower, where $w_n$ are the weights. In practice, to find $P_n(x)$, first define
\ba
I_n(m)&=&\int_m^\infty dx\, n^\prime(x) x^n=n(m) m^n- n m^{n-1}\ln(1-e^{-m})+\sum_{j=0}^{n-2} \frac{n! m^j}{j!} {\rm Li}_{n-j}\left(e^{-m}\right)\,.\nonumber
\ea
Constructing polynomials as $P_n(x)=x^n+a_{n-1}x^{n-1}+\ldots+a_1 x+ a_0$, the orthogonality (\ref{ortho}) can be written as
\ba
\left(\begin{array}{cccc}
  I_{n} & I_{n-1} & \ldots & I_0\\
  I_{n+1} & I_{n} & \ldots & I_1\\
  \ldots\\
  I_{2n} & I_{2n-1} &\ldots I_{n}
\end{array}\right)
\left(\begin{array}{c}
  a_{n}\\
  a_{n-1}\\
  \ldots\\
  a_{0}
\end{array}
\right)=-
\left(\begin{array}{c}
  I_{n+1}\\
  I_{n+2}\\
  \ldots\\
  I_{2n+1}
\end{array}
\right)\,,\nonumber
\ea
which leads to the coefficients $a_n$ after simple matrix inversion. From the coefficients $a_n$, one obtains the roots $x_n$ of the polynomial $P_{K+1}(x)$ as the eigenvalues of the companion matrix
\be
C=\left(\begin{array}{ccccc}
  0 & 0 & \ldots & 0 & -a_0\\
  1 & 0 & \ldots & 0 & -a_1\\
  0 & 1 & \ldots & 0 & - a_2\\
  \ldots\\
  0 & 0 & \ldots & 1 & -a_{K}
\end{array}
\right)\,.
\ee

Once the roots have been obtained, the weights $w_n$ are found from requiring (\ref{quadrature}) with $f(x)=\left\{1,x,x^2,\ldots, x^{K}\right\}$, requiring one more matrix inversion. This can be achieved efficiently using standard numerical matrix inversion packages.

\subsection{Shear viscosity in the 4d O(N) model}

As a sanity-check on the methodology and numerics, it is prudent to compare the results found as part of this work to the previous work in Ref.~\cite{Aarts:2004sd} for the O(N) model in 3+1 dimensions. There are several modifications in the formulae presented in this work when considering three spatial dimensions, which will be pointed out in this section.

First, the O(N) model in $4-2\epsilon$ dimensions at large N requires non-perturbative renormalization of the coupling constant which in $\overline{\rm MS}$ is given by
\be
\label{4dren}
\frac{1}{\lambda}_0=\frac{1}{\lambda}-\frac{1}{4\pi\epsilon}\,.
\ee
As a consequence, the theory possesses a Landau pole at large N, effectively rendering the theory ill defined at strong coupling. 
In terms of the renormalized coupling $\lambda$, the saddle point condition (\ref{saddle}) gets replaced by
\be
m=\frac{8 \sum_{n=1}^\infty \frac{K_1\left(\beta m n\right)}{n}}{\frac{4\pi^2}{\lambda}+1+\ln \frac{\bar\mu^2}{m^2}}\,,
\label{saddle3d}
\ee
and the entropy density for $d=4$ reads
\be
s=-\frac{N \beta}{\pi^2}\int_m^\infty dE E (E^2-m^2)^{\frac{3}{2}}\,,
\ee
cf. Eq.~(\ref{sr0}) and Ref.~\cite{Romatschke:2019gck}. (Note that (\ref{saddle3d}) possesses two solutions, one of which can be recognized as unphysical as long as the coupling $\lambda\ll 1$.) 

The Kubo formula for the $T^{xy}$ correlator is the same for d=3 and d=4, cf. Eq.~(\ref{grclassical}), so (\ref{Kubo}) holds for both cases. Also, expressions such as the thermal width (\ref{imsigma}) are formally the same, except that $\int_{\bf k}\equiv \int \frac{d^3{\bf k}}{(2\pi)^3}$ for d=4.
The vacuum contribution of the polarization tensor (\ref{pi}) is divergent in d=4 in dimensional regularization, but $\Pi$ only appears in the auxiliary field propagator (\ref{propagators}) in conjunction with $\frac{1}{\lambda_0}$, so that (\ref{4dren}) renders this combination finite. Effectively, this means there is a additional contribution $\frac{1}{8\pi^2}\ln\frac{\bar\mu^2}{m^2}$ to the real part of $\Pi_V$ not present for d=3.

\begin{figure*}[h]
  \includegraphics[width=.8\linewidth]{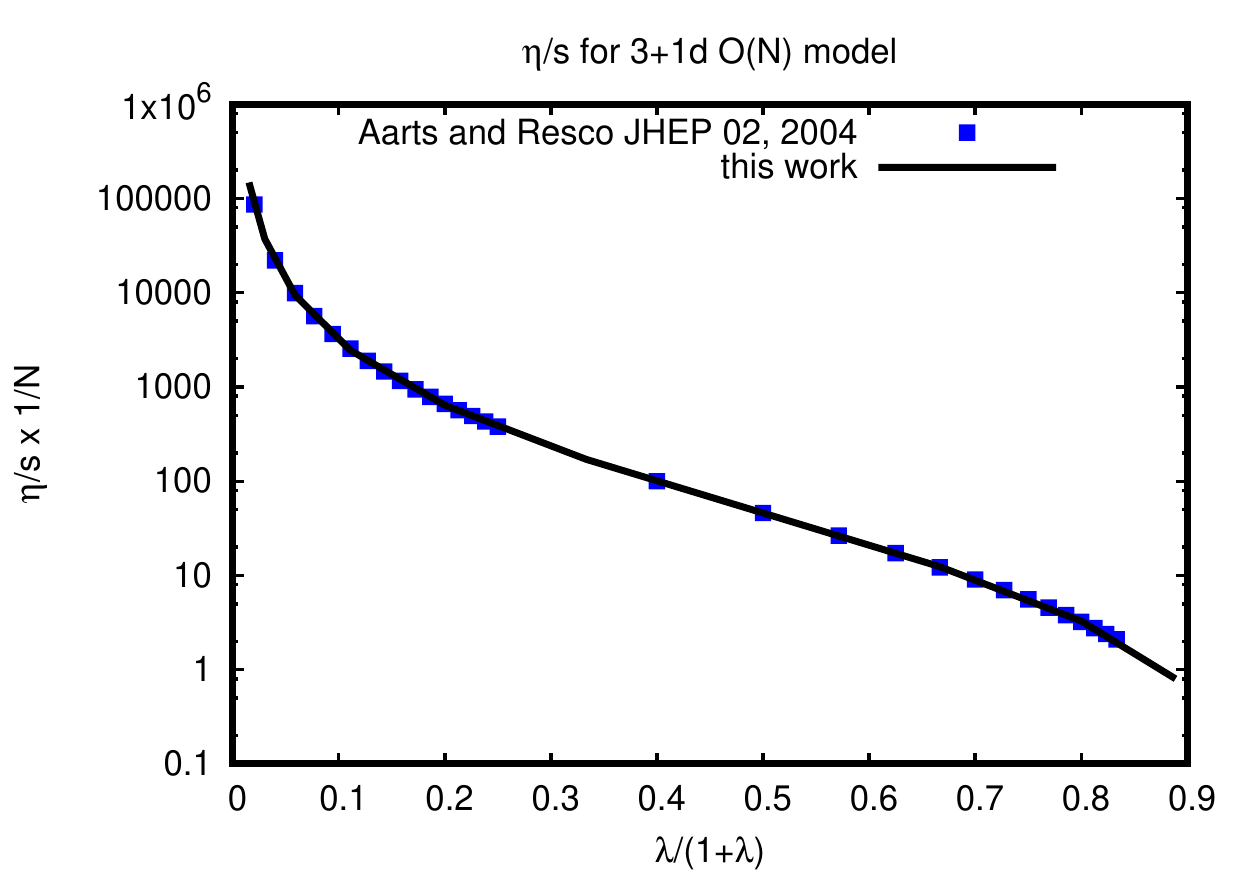}
  \caption{\label{fig2} Shear viscosity over entropy density as a function of $\lambda(\bar\mu=T)$ in the 3+1d O(N) model with quartic self-interaction. Horizontal axis is compactified in order to fit values $\lambda \in [0,\infty)$. For comparison, results from Aarts and Resco in Ref.~\cite{Aarts:2004sd} are shown. One finds reasonable agreement between the numerical results from Ref.~\cite{Aarts:2004sd} and this work.}
    \end{figure*}

The quadrature rules (\ref{ortho}) are modified in d=4 to read
\be
\label{ortho4}
\int_m^\infty dx\, \sqrt{x^2-m^2}n^\prime(x) P_i(x) P_j(x)=\delta_{ij}\,,
\ee
and are otherwise constructed analogously to the d=3 case.

The vertex resummation is formally identical to (\ref{vertex2}) except for the angular weight (\ref{2dweight}) which for d=4 becomes
\be
\label{3dweight}
f(x)=\frac{1}{2}\left(3 x^2-1\right)=P_2(x)\,,
\ee
also known as the second Legendre polynomial. As a consequence, the shear-viscosity over entropy density ratio is modified from (\ref{fulleta}) to read in d=4
\be
\label{fulleta3d}
\frac{\eta}{s}=\frac{1}{5}\frac{\int_m^\infty dE \frac{F(E_k)}{\beta {\rm Im}\, \Sigma(k)}k^3 n^\prime(E)}{-\int_m^\infty dE E k^3 n^\prime(E)}\,.
\ee

Evaluating (\ref{fulleta3d}) using the same numerical techniques as for d=3, one finds the results shown in Fig.~\ref{fig2}. As can be seen from this figure, there is reasonable agreement between the present work and the published results from Ref.~\cite{Aarts:2004sd}, serving as a sanity-check of both the algorithm and the numerical technique used here.

\bibliography{bibi}

\end{document}